\documentclass[prb,aps,twocolumn,superscriptaddress,showpacs]{revtex4-1}
\usepackage{graphicx}
\usepackage{amsmath}
\usepackage{epsfig}
\usepackage{multirow}
\usepackage{array}
\usepackage{inputenc}
\usepackage{color}
\usepackage{tabularx}
\usepackage{textcomp}
\usepackage{ulem}
\usepackage[colorlinks=true,linkcolor=black, citecolor=blue, urlcolor=blue, 
    unicode=true]{hyperref}

\begin{document}

\title{Kitaev interactions in the Co honeycomb antiferromagnets Na$_3$Co$_2$SbO$_6$ and Na$_2$Co$_2$TeO$_6$}

\author{M. Songvilay}
\affiliation{School of Physics and Astronomy and Centre for Science at Extreme Conditions, University of Edinburgh, Edinburgh EH9 3FD, UK}
\affiliation{Institut N\'eel, CNRS \& Univ. Grenoble Alpes, 38000 Grenoble, France}
\author{J. Robert}
\affiliation{Institut N\'eel, CNRS \& Univ. Grenoble Alpes, 38000 Grenoble, France}
\author{S. Petit}
\affiliation{Laboratoire L\'{e}on Brillouin, CEA-CNRS UMR 12, 91191 Gif-Sur-Yvette Cedex, France}
\author{J. A. Rodriguez-Rivera}
\affiliation{NIST Center for Neutron Research, National Institute of Standards and Technology, 100 Bureau Drive, Gaithersburg, Maryland,20899, USA}
\affiliation{Department of Materials Science, University of Maryland, College Park, Maryland 20742, USA}
\author{W. D. Ratcliff}
\affiliation{NIST Center for Neutron Research, National Institute of Standards and Technology, 100 Bureau Drive, Gaithersburg, Maryland, 20899, USA}
\author{F. Damay}
\affiliation{Laboratoire L\'{e}on Brillouin, CEA-CNRS UMR 12, 91191 Gif-Sur-Yvette Cedex, France}
\author{V. Bal\'edent}
\affiliation{Laboratoire de Physique des Solides, CNRS, Universit\'e Paris-Sud, Universit\'e Paris-Saclay, 91405 Orsay Cedex, France}
\author{M. Jim\'enez-Ruiz}
\affiliation{Institut Laue-Langevin, 71 avenue des Martyrs, 38000 Grenoble, France}
\author{P. Lejay}
\affiliation{Institut N\'eel, CNRS \& Univ. Grenoble Alpes, 38000 Grenoble, France}
\author{E. Pachoud}
\affiliation{Institut N\'eel, CNRS \& Univ. Grenoble Alpes, 38000 Grenoble, France}
\author{A. Hadj-Azzem}
\affiliation{Institut N\'eel, CNRS \& Univ. Grenoble Alpes, 38000 Grenoble, France}
\author{V. Simonet}
\affiliation{Institut N\'eel, CNRS \& Univ. Grenoble Alpes, 38000 Grenoble, France}
\author{C. Stock}
\affiliation{School of Physics and Astronomy and Centre for Science at Extreme Conditions, University of Edinburgh, Edinburgh EH9 3FD, UK}
\date{\today}

\begin{abstract}
Co$^{2+}$ ions in an octahedral crystal field, stabilise a $j_{\rm eff}=1/2$ ground state with an orbital degree of freedom and have been recently put forward for realising Kitaev interactions, a prediction we have tested by investigating spin dynamics in two cobalt honeycomb lattice compounds, Na$_2$Co$_2$TeO$_6$ and Na$_3$Co$_2$SbO$_6$, using inelastic neutron scattering. We used linear spin wave theory to show that the magnetic spectra can be reproduced with a spin Hamiltonian including a dominant Kitaev nearest-neighbour interaction, weaker Heisenberg interactions up to the third neighbour and bond-dependent off-diagonal exchange interactions. Beyond the Kitaev interaction that alone would induce a quantum spin liquid state, the presence of these additional couplings is responsible for the zigzag-type long-range magnetic ordering observed at low temperature in both compounds. These results provide evidence for the realization of Kitaev-type coupling in cobalt-based materials, despite hosting a weaker spin-orbit coupling than their $4d$ and $5d$ counterparts.

\end{abstract}

\pacs{}

\maketitle


\section{Introduction}
The search for quantum spin liquids in frustrated honeycomb lattice materials  \cite{Savary2017,Meng2010,Fouet2001} has been recently renewed by the theoretical model proposed by A. Kitaev describing spins interacting on a honeycomb lattice through strongly anisotropic bond-directional couplings (hereafter called Kitaev interactions) \cite{Kitaev2006}. Such model is exactly solvable and predicts a quantum spin liquid ground state with exotic anyonic excitations, appealing for quantum information technologies \cite{Nayak2008}. Following this theoretical model, Jackeli and Khaliullin paved the way to the realization of Kitaev interactions in real materials, proposing its achievement through a strong interplay between spin-orbit coupling and electronic correlations \cite{Jackeli2009}. In their proposal, the large spin-orbit coupling present in the $4d$ and $5d$ transition metals, such as iridium and ruthenium, allows a spin-orbital entangled $j_{\rm eff} = 1/2$ Kramers doublet ground state 
and a few compounds within the A$_2$IrO$_3$ (A = Na, Li) family \cite{Chaloupka2013,Choi2012,Chun2015,Gretarsson2013,Kitagawa2018,Singh2010,Takayama2015}  and $\alpha$-RuCl$_3$ \cite{Banerjee2016,Banerjee2017,Banerjee2018,Do2017,Jansa2018,Kasahara2018,Sandiland2015,Sandiland2016,Sears2015,Sears2017,Winter2018,Wu2018,Rau2014} have been identified as potential candidates for exhibiting Kitaev physics.
It has been more recently suggested that the $d^7$ Co$^{2+}$ ions can also be used to create a $j_{\rm eff}=1/2$ system with an orbital degree of freedom, and therefore may also realize Kitaev interactions \cite{Sano2018,Liu2018,Liu2020,Plumb2014}, despite a weaker spin-orbit coupling.  We therefore propose in this letter to re-examine two cobalt honeycomb lattice compounds in the context of the recently proposed Kitaev model.

\begin{figure}
 \includegraphics[scale=0.45]{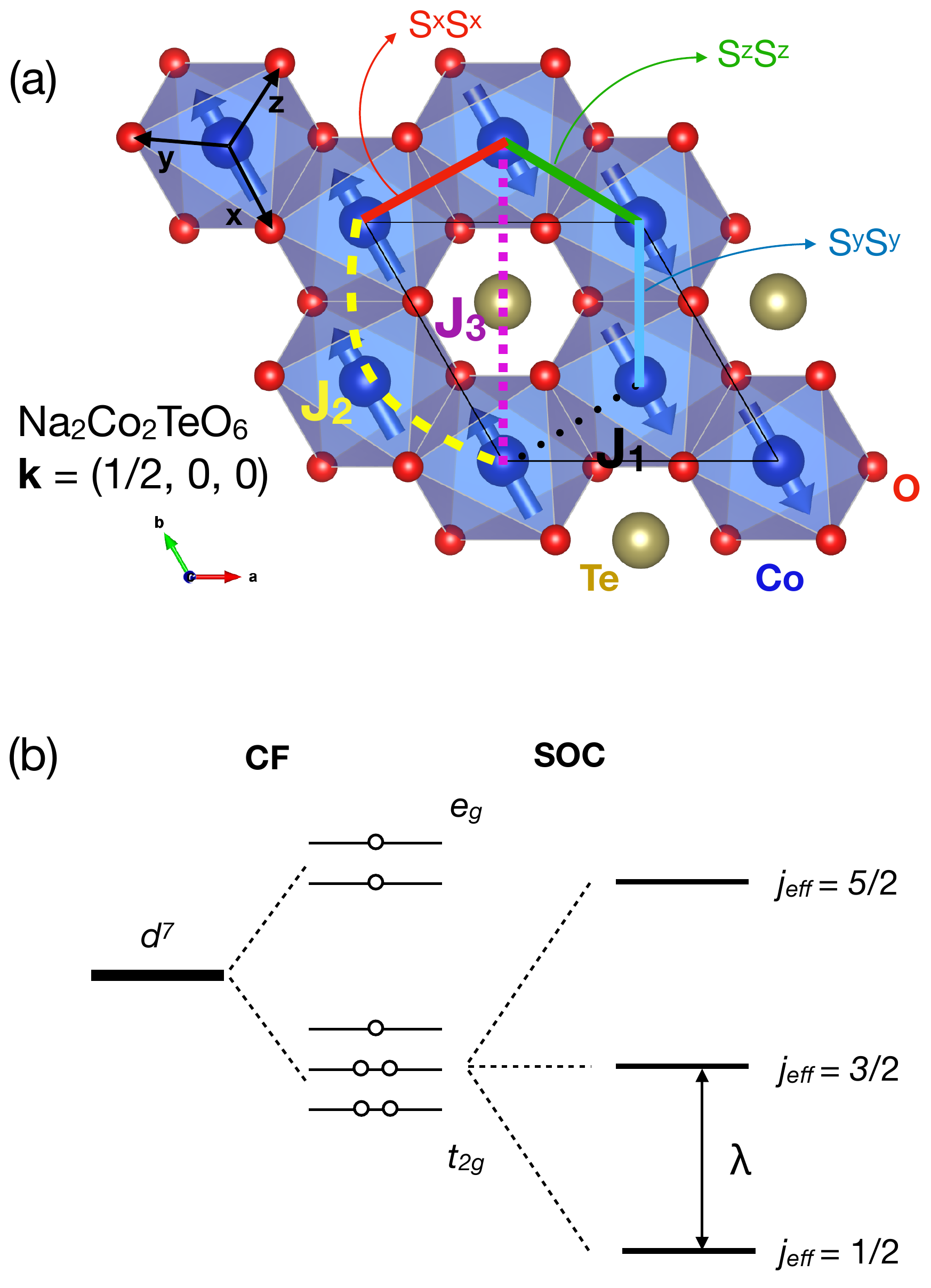}
 \caption{\label{fig:structure} $(a)$ Honeycomb layer with the Co$^{2+}$ ions in edge-sharing octahedra. The zigzag magnetic structure within the plane stabilized in Na$_2$Co$_2$TeO$_6$ is shown (blue arrows), as well as the first to third nearest-neighbour Heisenberg interactions. The red, blue and green lines indicate the three nearest-neighbour bonds associated to the local $x$, $y$, and $z$ axes in the Kitaev model  $(b)$ Electronic configuration of $3d^7$ Co$^{2+}$ in an octahedral environment and representation of the $j_{\rm eff}$ = 1/2 ground state, and excited $j_{\rm eff}$ = 3/2 and $j_{\rm eff}$ = 5/2 manifolds under the action of crystal field and spin-orbit coupling.}
\end{figure}

In this framework, the recently synthesized and characterized Na$_2$Co$_2$TeO$_6$ and  Na$_3$Co$_2$SbO$_6$  compounds \cite{Viciu2007} have been identified as good candidates for testing this proposal \cite{Liu2018}. Both systems host a honeycomb lattice of interacting Co$^{2+}$ ions in edge-sharing oxygen octahedra. Na$_2$Co$_2$TeO$_6$ and Na$_3$Co$_2$SbO$_6$ belong to a wider family of delafossite-related compounds where the sodium site can be replaced by Li, Ag, Cu or Ni \cite {Skakle1997,Zvereva2016,Roudebush2013}. They crystallise in the hexagonal $P$6$_3$22 and monoclinic $C2/m$ space group, respectively. While in the Sb compound the layers of Co ions are separated by ordered Na ions, the Na distribution in the Te counterpart is highly disordered \cite{Viciu2007}. Both systems undergo a magnetic transition around 17~K for the Te compound and between 4 and 8~K for the Sb compound, into a zigzag magnetic order \cite{Lefrancois2016,Bera2017,Wong2016,Yan2019} as illustrated in Figure \ref{fig:structure}. In the case of the Te compound, it is not fully long-range ordered and characterized by anisotropic correlation lengths. The orientation of the moments are mainly along the $b$ axis but a small component along $c$ is not excluded \cite{Lefrancois2016}. In the Sb compound, the moments have been reported to point along the $c$ or the $b$ axis of the monoclinic space group \cite{Lefrancois2016,Bera2017,Wong2016,Yan2019}. This magnetic structure is predicted in both the Kitaev-Heisenberg model in proximity of the spin liquid phase \cite{Chaloupka2013} and in the Kitaev-Heisenberg model with next-nearest-neighbor couplings, $J_2$ and $J_3$ \cite{Kimchi2011,Winter2017}.

In this work, we performed inelastic neutron scattering measurements on polycrystalline samples of these materials to investigate the magnetic excitations in their ordered phase. We show, using linear spin wave calculations, that both systems can be modeled with a Kitaev-Heisenberg-$J_2$-$J_3$ Hamiltonian characterized by dominant Kitaev interactions, like the well-known candidate $\alpha$-RuCl$_3$ \cite{Winter2017}, and we compare this model against the pure Heisenberg $J_1$-$J_2$-$J_3$ model which also encompasses a zigzag ordered groundstate.

In these cobaltate systems, the $3d^7$ Co$^{2+}$ ions in an octahedral environment have a predominantly $t_{2g}^5$ $e_g^2$ configuration with $S$ = 3/2 and an effective $\tilde{l}$ = 1 moments, and form a $j_{\rm eff}=1/2$ Kramers doublet ground state (Fig. \ref{fig:structure} $(b)$) \cite{AbragamBleaney, Sarte2018, Sarte2018bis, Sarte2019}. Owing to the electronic configuration, anisotropic interactions are mediated through the oxygen atoms in a 90$^{\circ}$ bonding geometry, including exchange processes via $t_{2g}$-$e_g$ and $e_g$-$e_g$ channels. These channels do not exist in the well-studied $4d$ and $5d$ systems, and it has been shown that these can lead to a cancellation of the Heisenberg couplings in favour of a Kitaev term $K$ \cite{Liu2018,Sano2018}. Cobaltates thus provide an alternative way to reach the $K>>J_1$ spin-liquid ground state.

\section{Synthesis and Experimental details}
Polycrystalline samples of Na$_2$Co$_2$TeO$_6$ and Na$_3$Co$_2$SbO$_6$ were prepared following the methods described in \cite{Viciu2007,Lefrancois2016,Wong2016}. Neutron spectroscopy was performed to study the magnetic excitations at low energy, using the cold triple-axis spectrometer MACS (Multi Axis Crystal Spectrometer) \cite{MACS} (NIST, Gaithersburg) with a fixed E$_f$~=~5~meV and a Be filter between the sample and the analyser to avoid higher order contamination. After subtracting a constant background from the raw data, the neutron scattering intensity was converted to absolute units using the measured incoherent elastic scattering from the sample following the procedure described in Ref. \cite{Hammar1999}. Further measurements using the BT4 thermal triple-axis spectrometer (NIST, Gaithersburg) were carried out to map out the excitations between the $j_{\rm eff}$ = $\frac{1}{2}$ groundstate and the $j_{\rm eff}$ = $\frac{3}{2}$ excited states and extract the spin-orbit coupling $\lambda$. The measurements were performed with a fixed E$_f$~=~14.7~meV and a 40' incident collimation. A pyrolytic graphite (PG) filter was placed between the sample and the analyser to avoid higher order contamination. Complementary measurements were carried out up to 150 meV on the LAGRANGE (Large Area GRaphite ANalyser for Genuine Excitations) spectrometer at the Institut Laue Langevin (ILL, France), using the Si111, Si311 and Cu220 monochromators to access different energy ranges between 0 and 150 meV (see Supplemental Information). 
Spin wave calculations were performed in the linear approximation using the Holstein-Primakov formalism \cite{Petit2011}.
Additional neutron powder diffraction on Na$_3$Co$_2$SbO$_6$ was carried out using the G4.1 diffractometer (LLB, Saclay) with $\lambda$~=~2.43~\AA~and powder x-ray diffraction was performed on the same compound at the Laboratoire de Physique des Solides (Orsay). The powder was loaded into a sealed 0.5 mm diameter capillary and put on a rotating sample holder. During the 30 minutes acquisition, the sample performed a 360 degrees rotation in order to average any possible texture. The X-ray generator was a Mo rotating anode, with a wavelength of 0.71~\AA (see Supplemental Information).


\section{Results and analysis}
\subsection{Neutron inelastic scattering}
\begin{figure}
\includegraphics[scale=0.35]{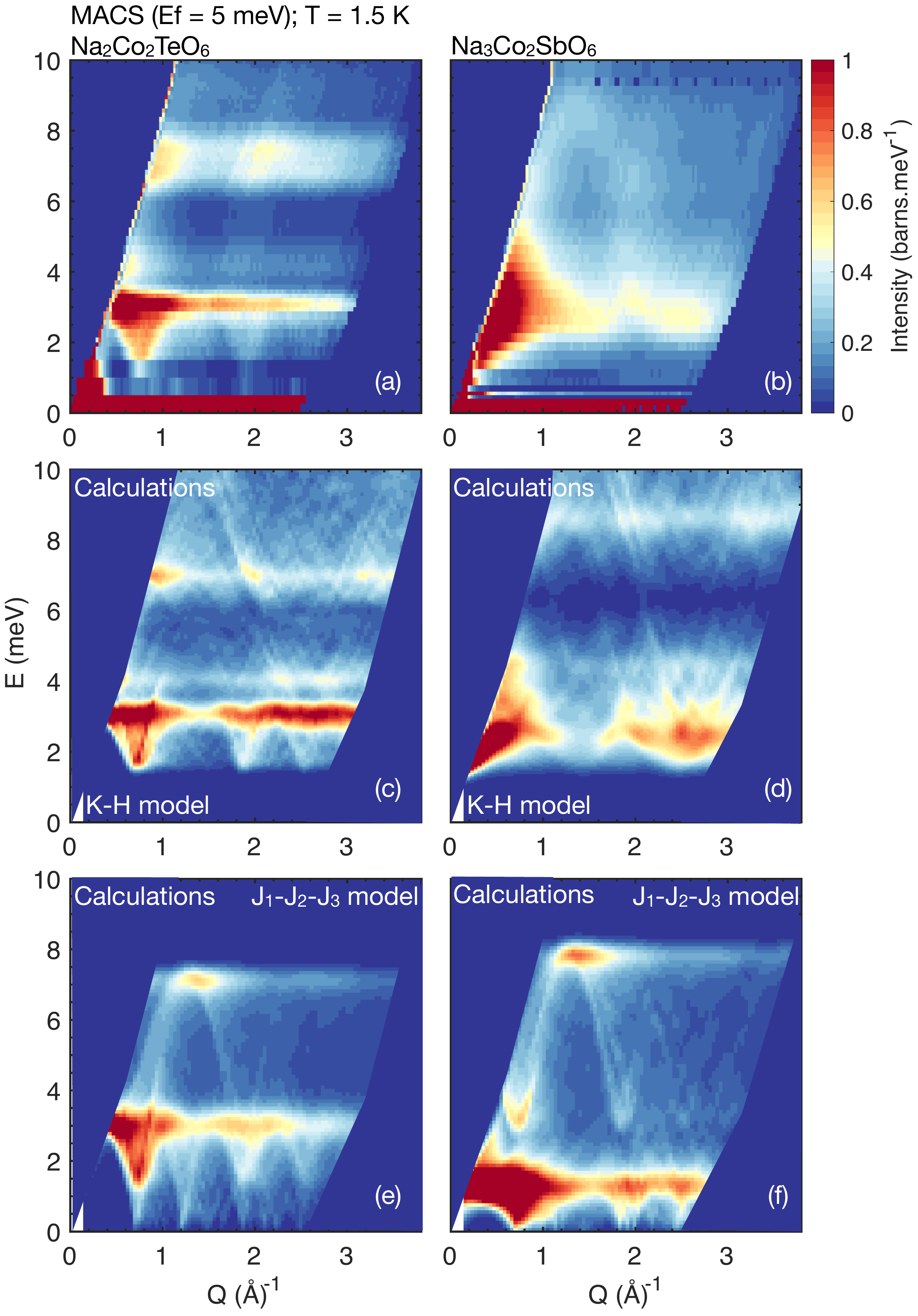}
 \caption{\label{fig:macs} $(a)$, $(b)$ S($\mathbf{Q}$, $\omega$) at T = 1.5 K measured on MACS in Na$_2$Co$_2$TeO$_6$ and Na$_3$Co$_2$SbO$_6$, respectively. $(c)$, $(d)$ Spinwave calculations using the K-H Hamiltonian with the parameter values reported in Table \ref{tab:couplings}. $(e)$, $(f)$ Spinwave calculations using the $J_1$-$J_2$-$J_3$ Heisenberg model as described in the text and in Figure \ref{fig:XXZparam}. }
\end{figure}

We first discuss the spin dynamics in both compounds in their ordered phase. Figures \ref{fig:macs}$(a)$ and $(b)$ show the magnetic excitation spectra of Na$_2$Co$_2$TeO$_6$ and Na$_3$Co$_2$SbO$_6$ respectively, measured at T~=~1.5~K on MACS. Remarkably, while both compounds order with the same zig-zag magnetic structures below T$_N$ \cite{Lefrancois2016,Wong2016,Yan2019}, their excitation spectra show rather different features, in an energy range between 0 and 10 meV. For the Te compound, the spectrum consists in gapped dispersive modes merging into a flat mode at 3~meV. The first dispersive branches arise from Q~$\approx$~0.75~\AA$^{-1}$. Two other flat modes in momentum transfer are observed, a weak one around 4.5~meV and a broad one at 7~meV, while weak signal is still noticeable up to 10 meV. In contrast, the Sb compound displays a dispersive signal up to 9 meV with an intense part at low Q between 1 and 5 meV  and an additional flatter mode at 3 meV with a small kink around Q~=~2~\AA$^{-1}$.

Further measurements at higher energy were also carried out using the thermal triple-axis spectrometer BT4 (Figure \ref{fig:crystal_field}) and the LAGRANGE spectrometer (see Supplemental Information). Figure \ref{fig:crystal_field} $(a)$ shows the momentum dependence of a weakly dispersive excitation mode observed around 30 meV in Na$_3$Co$_2$SbO$_6$, for which the intensity follows the expected momentum dependence of the magnetic form factor. We can therefore attribute this mode to spin-orbit excitations between the $j_{\rm eff}$~=~1/2 and the $j_{\rm eff}$~=~3/2 manifolds. Constant-Q scans were performed in an energy range between 10 and 40 meV for both compounds (Figure \ref{fig:crystal_field} $(d)$ and $(e)$) and a fit to Gaussian functions yields energy values of 27.1(2) meV for the Sb compound, and 21.6(1) meV for the Te compound. These energy scales are similar to those reported in other Co-based materials, where the spin-orbit transitions were found between 16 and 34 meV \cite{Wallington2015, Sarte2018, Sarte2018bis, Sarte2019}.

\begin{figure}
 \includegraphics[scale=0.35]{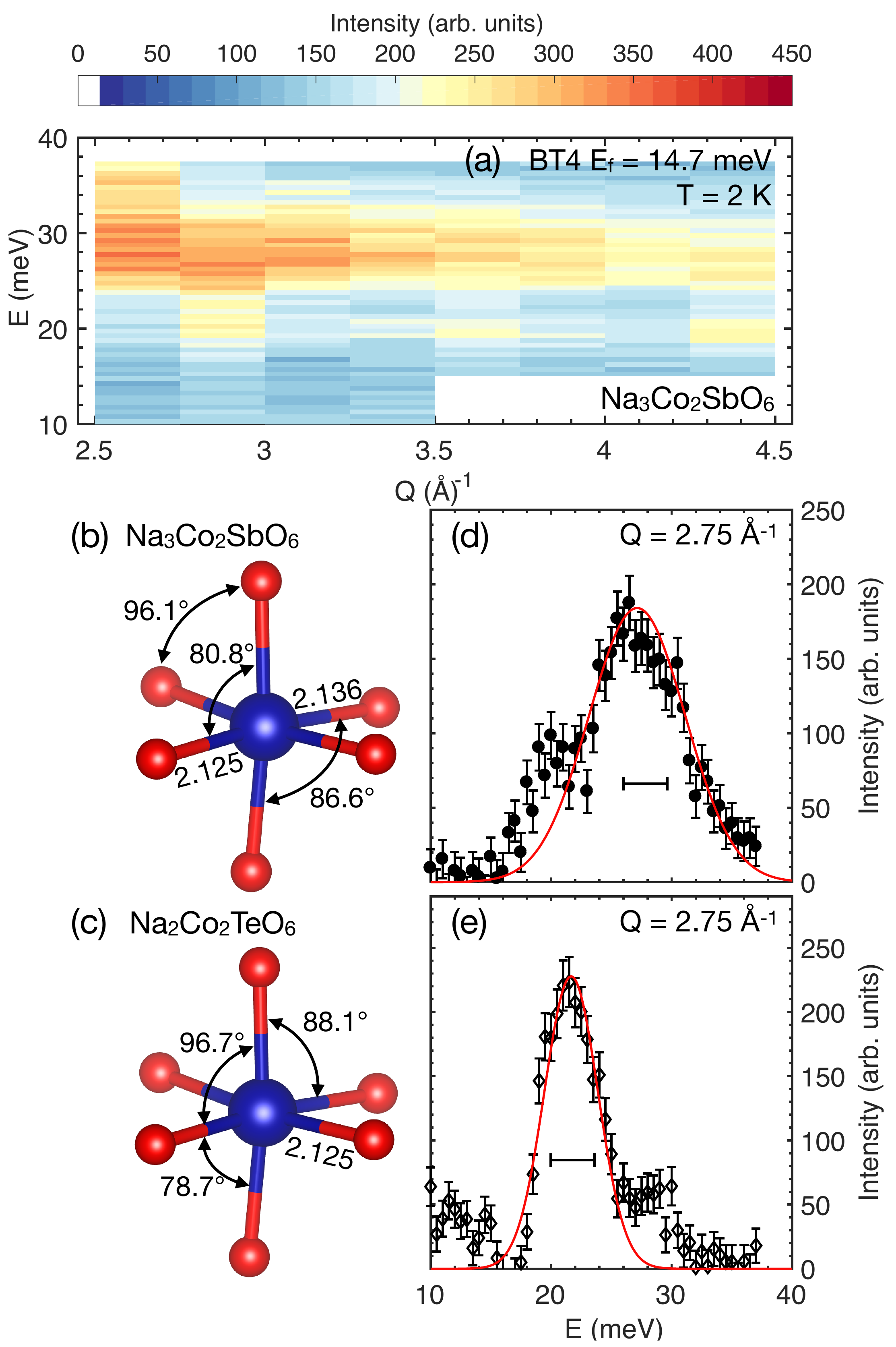}
 \caption{\label{fig:crystal_field} $(a)$ Colormap showing the Q dependence of the spin-orbit excitations between the $j_{\rm eff}$ = 1/2 and $j_{\rm eff}$ = 3/2 manifolds measured at T = 2 K for Na$_3$Co$_2$SbO$_6$. $(b)$-$(c)$ View of the slightly distorted CoO$_6$ octahedra in Na$_3$Co$_2$SbO$_6$ and Na$_2$Co$_2$TeO$_6$, respectively. $(d)$-$(e)$ Constant-Q scans showing the spin-orbit excitations between the $j_{\rm eff}$ = 1/2 and $j_{\rm eff}$ = 3/2 manifolds in Na$_3$Co$_2$SbO$_6$ and Na$_2$Co$_2$TeO$_6$, respectively. The straight lines indicate the instrumental resolution and the red lines are gaussian fits. The additional peaks are phonon modes}
\end{figure}

\subsection{Sum rules}

 In order to confirm the assignment of the observed magnetic excitations at low energy to excitations within the $j_{\rm eff}$ = 1/2 manifold \cite{Cowley2013}, we apply the total moment sum rule by integrating the normalised spectral weight from the low energy spectra. First, the inelastic intensity $I(\textbf{Q},\omega)$ was normalised to absolute units and converted to the dynamic structure factor $S(\textbf{Q},\omega)$ following the relation: 

\[AI(\textbf{Q},\omega) = 2\left( \frac{\gamma r_0}{2} \right)^2 g^2 |f(\textbf{Q})|^2 S(\textbf{Q},\omega)   \]

where $A$ is the absolute calibration constant calculated using Co and Na as internal incoherent standards \cite{Sears1992}, $ \left( \frac{\gamma r_0}{2} \right)^2$ equals 73 mb sr$^{-1}$, $g$ is the Land\'e factor equal to 4.33 for Co$^{2+}$ \cite{AbragamBleaney,Sarte2018bis} and $f(\textbf{Q})$ is the magnetic structure factor calculated for Co$^{2+}$.

\begin{figure}
 \includegraphics[scale=0.35]{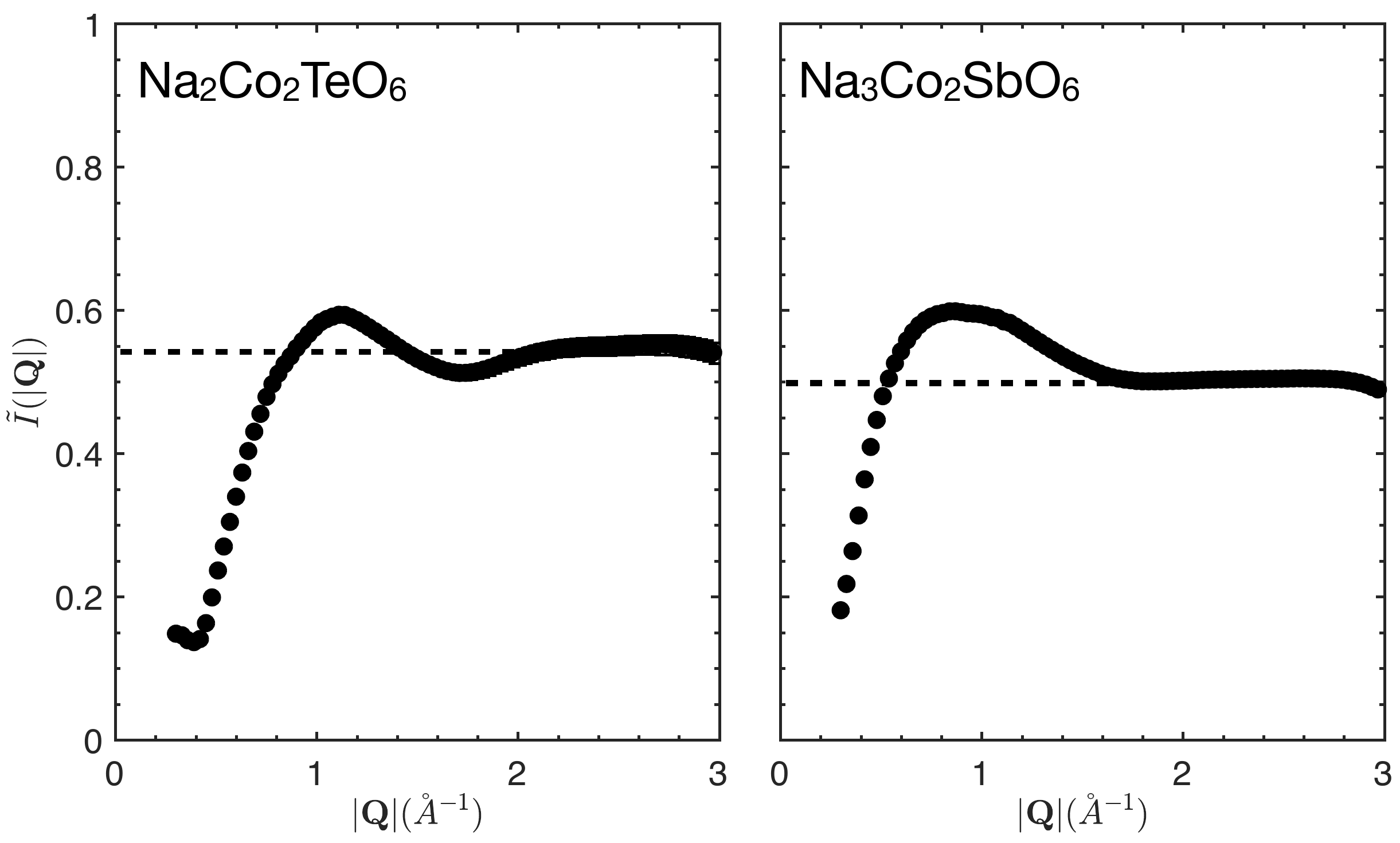}
 \caption{\label{fig:sumrule} Momentum dependence of the energy integrated inelastic intensity of Na$_2$Co$_2$TeO$_6$ (left) and Na$_3$Co$_2$SbO$_6$ (right).}
\end{figure}

The dynamic structure factor then obeys the total moment sum rule:

 \[\frac{\int |\textbf{Q}|^2 \left(  \int S(\textbf{Q},\omega) \hbar d\omega \right) d|\textbf{Q}| }{\int d|\textbf{Q}| |\textbf{Q}|^2} = Nj(j+1) \]
 
with $N$ the number of Co ions in a unit cell.
Figure \ref{fig:sumrule} displays, for both compounds, the Q dependence of the total integrated inelastic intensity given by 

\[ \tilde{I}|(\mathbf{Q})| = \frac{\int_0 ^{|\mathbf{Q}|} |\textbf{Q}|^2 \left(  \int S(\textbf{Q},\omega)  \hbar d\omega \right)d|\textbf{Q}|}{\int_0 ^{|\mathbf{Q}|} d|\textbf{Q}||\textbf{Q}|^2}  \]

 The results show that $\tilde{I}(|\mathbf{Q}|)$ saturates at 0.55(15) for Na$_2$Co$_2$TeO$_6$ and 0.50(15) for Na$_3$Co$_2$SbO$_6$.
Adding the elastic contributions to the inelastic integrated signal, to extract the total moment sum rule, yields the values tabulated in table \ref{tab:sumrule}. For the elastic contribution, the averaged refined ordered moment of $\left( \frac{(2.5 + 2.95)}{2} \right)^2 \left( \frac{1}{g\mu_B} \right) ^2$ was used for Na$_2$Co$_2$TeO$_6$ \cite{Lefrancois2016} and for Na$_3$Co$_2$SbO$_6$  the elastic magnetic Bragg peak intensity was integrated yielding the contribution $\left(\frac{1.79}{g\mu_B}\right)^2$.
The total value extracted in both cases is within error of the expected value of $j_{\rm eff}$ = 1/2, which is $j(j+1) $= 0.75 and confirms the assignment of the low-energy spectrum to excitations within the ground-state $j_{\rm eff}$ = 1/2 manifold, thus validating the $j_{\rm eff}$~=~1/2 picture in these materials.
Combined together, the low and higher energy parts of the excitation spectrum show that the $j_{\rm eff}$~=~1/2 and the $j_{\rm eff}$~=~3/2 manifolds seem rather well separated in energy, with excitations within the $j_{\rm eff}$~=~1/2 manifold extending below 15 meV and the spin-orbit excitation between 20 and 30 meV in both compounds. 

\begin{table}[t]
\caption{\label{tab:sumrule} Distribution of the spectral weight, extracted from the elastic and inelastic contributions to the total intensity in Na$_2$Co$_2$TeO$_6$ and Na$_3$Co$_2$SbO$_6$.}
\begin{ruledtabular}
\begin{tabular}{llll}
    &   Na$_2$Co$_2$TeO$_6$ & Na$_3$Co$_2$SbO$_6$  \\
 \hline
Elastic & 0.40 from \cite{Lefrancois2016} & 0.17(2) \\
Inelastic  & 0.55(15) & 0.50(15)  \\
total & 0.95(20) & 0.67(20) \\
\end{tabular}
\end{ruledtabular}
\end{table}

\subsection{Linear spin-wave calculations}

In order to extract the magnetic interactions in each system and model the inelastic spectra, we performed linear spin wave calculations. Following the recent analysis on $d^7$ Co ions with unquenched orbital contribution realising bond-dependent interactions \cite{Liu2018,Sano2018}, the inelastic data was modeled using the recently proposed Kitaev-Heisenberg Hamiltonian (K-H model) \cite{Liu2020,Maksimov2020}. The considered Hamiltonian is the sum of six terms $J_1$, $J_2$, $J_3$, $K$, $\Gamma$, $\Gamma'$ (as illustrated in Figure \ref{fig:structure}), authorized by symmetry when including slight distortions of the Co$^{2+}$ octahedra: 

\begin{eqnarray*}
H_{K-H} = & &\sum_{n=1}^3 J_n \sum_{i,j}  \mathbf{S}_i \cdot \mathbf{S}_j + \sum_{i,j} K S^{\gamma}_i S^{\gamma}_j  \\ 
&+  &\sum_{i,j} \Gamma (S^{\alpha}_i S^{\beta}_j + S^{\beta}_i S^{\alpha}_j) \\
&+ &\sum_{i,j} \Gamma' (S^{\alpha}_i S^{\gamma}_j + S^{\gamma}_i S^{\alpha}_j + S^{\beta}_i S^{\gamma}_j  + S^{\gamma}_i S^{\beta}_j )
\end{eqnarray*}

Where $\{\alpha,\beta,\gamma\}$ denotes the three types of first neighbor bonds with $\{\alpha,\beta,\gamma\} = \{y, z, x\}, \{z, x, y\}, \{x, y, z\}$ for the X, Y and Z-bonds respectively, and $\Gamma$ and $\Gamma'$  are bond-dependent off-diagonal exchange interactions. The model includes Heisenberg first, second and third-neighbor interactions as illustrated in Figure 1, which are contained in the first summation over $n$. 

\begin{table}[t]
\caption{\label{tab:couplings} Set of interactions (in meV) reproducing the magnetic spectra for Na$_2$Co$_2$TeO$_6$ and Na$_3$Co$_2$SbO$_6$. All the parameters are inter-dependent and the sensitivity of the spectra to their variation is illustrated in figures \ref{fig:KHparam_Te} and \ref{fig:KHparam_Sb}.}
\begin{ruledtabular}
\begin{tabular}{lllllll}
  & $J_1$ & $J_2$ & $J_3$ & $K$ & $\Gamma$ & $\Gamma'$ \\
 \hline
 Na$_2$Co$_2$TeO$_6$ & -0.1(5) & 0.3(3) & 0.9(3) & -9.0(5) & 1.8(5) & 0.3(3)\\
 \hline
 Na$_3$Co$_2$SbO$_6$ & -2.0(5) & 0.0(2) & 0.8(2) & -9.0(10) & 0.3(3) & -0.8(2)\\
\end{tabular}
\end{ruledtabular}
\end{table}
 
\begin{figure}
 \includegraphics[scale=0.32]{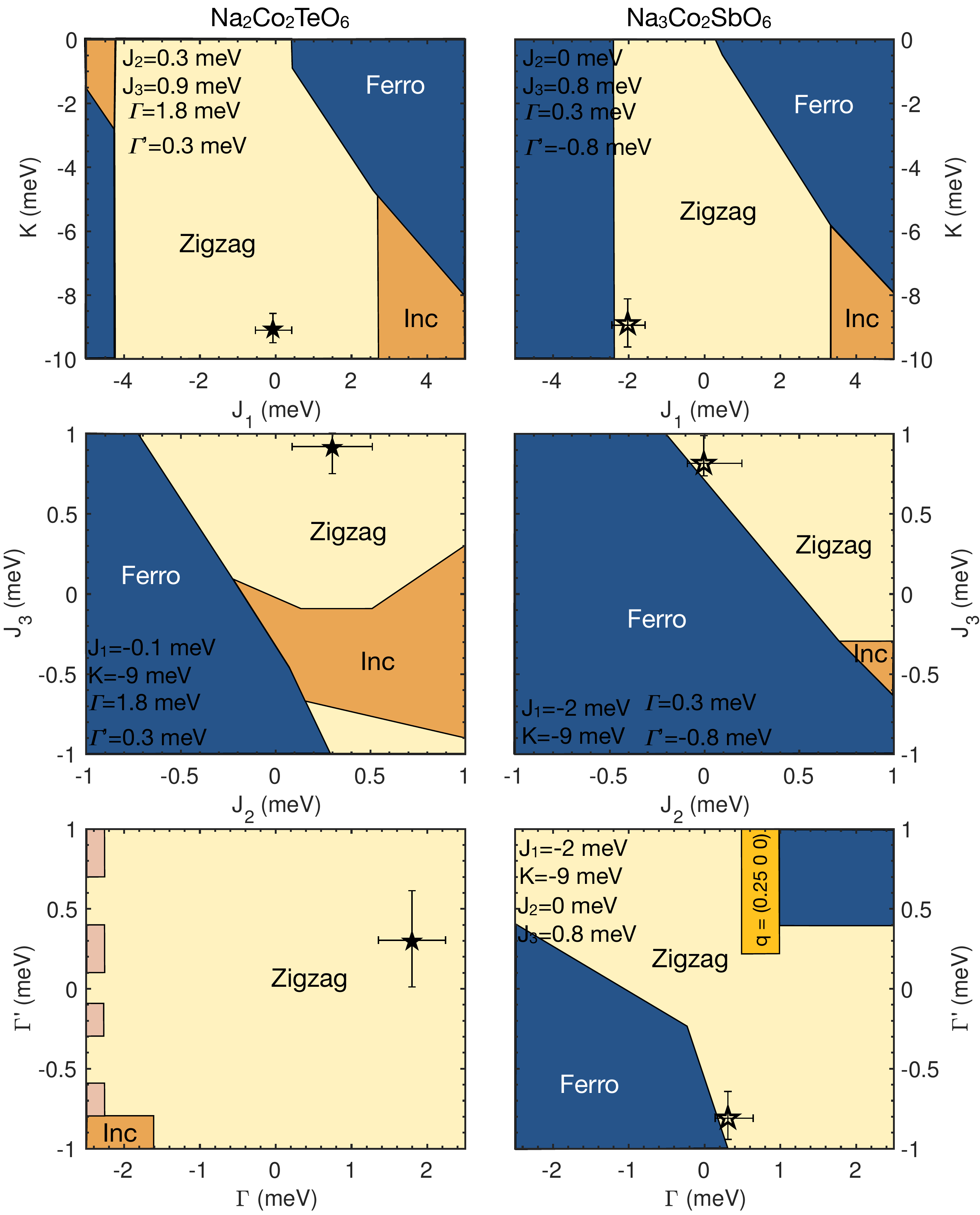}
 \caption{\label{fig:diagphase} Magnetic phase diagrams calculated for Na$_2$Co$_2$TeO$_6$ (left panels) and Na$_3$Co$_2$SbO$_6$ (right panels) for different pairs of parameters. The blue, yellow and orange regions correspond to ferromagnetic, zig-zag and incommensurate magnetic structures while the pink areas correspond to a degenerate groundstate. The black stars show the best-fit solution.}
\end{figure}

First, phase diagrams were calculated (Figure \ref{fig:diagphase}) in order to check the stability of the zigzag magnetic order as a function of the six parameters.  In order to best reproduce the experimental spectra, several sets of parameters which are compatible with the zigzag magnetic structure were systematically tested in the spin-wave analysis. The solutions that best reproduce the data are given in Table \ref{tab:couplings} and the chosen parameters are represented in the phase diagram in Figure \ref{fig:diagphase}, while the calculated spectra are shown in Figures \ref{fig:macs}$(c)$ and $(d)$ for Na$_2$Co$_2$TeO$_6$ and Na$_3$Co$_2$SbO$_6$, respectively. By varying each parameter and testing how each affects the spectra, we could determine a narrow region in the phase diagram (represented by the error bars in Figure \ref{fig:diagphase}) which gives the best agreement with the data. Moreover, the experimental data was also compared to a Heisenberg XXZ Hamiltonian, comprising a first, second and third-nearest-neighbour interactions. We describe these tests in details in the next two sections.

\subsubsection{The Kitaev-Heisenberg model}

\begin{figure*}
 \includegraphics[scale=0.55]{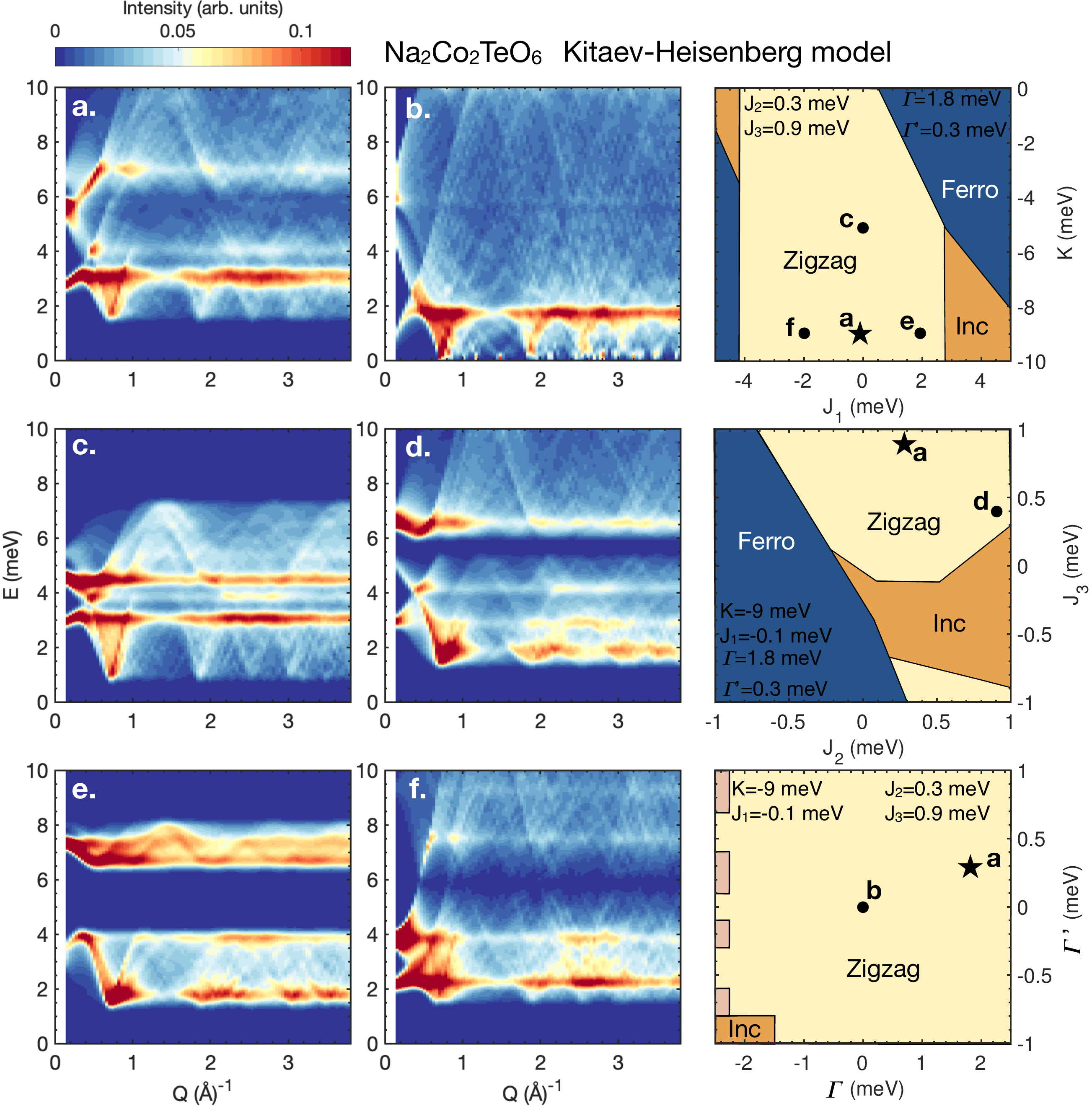}
 \caption{\label{fig:KHparam_Te} Dependence of the spin-wave calculation with respect to variation of the six parameters involved in the K-H model for Na$_2$Co$_2$TeO$_6$. The panel $(a)$ corresponds to the chosen solution, that best reproduces the experimental data. Each panel ($(a)$-$(f)$) corresponds to a set of parameters, which are represented in the phase diagram (right panels). The blue, yellow and orange regions correspond to ferromagnetic, zig-zag and incommensurate magnetic structures while the pink areas correspond to a degenerate groundstate.}
\end{figure*} 

\begin{figure*}
 \includegraphics[scale=0.55]{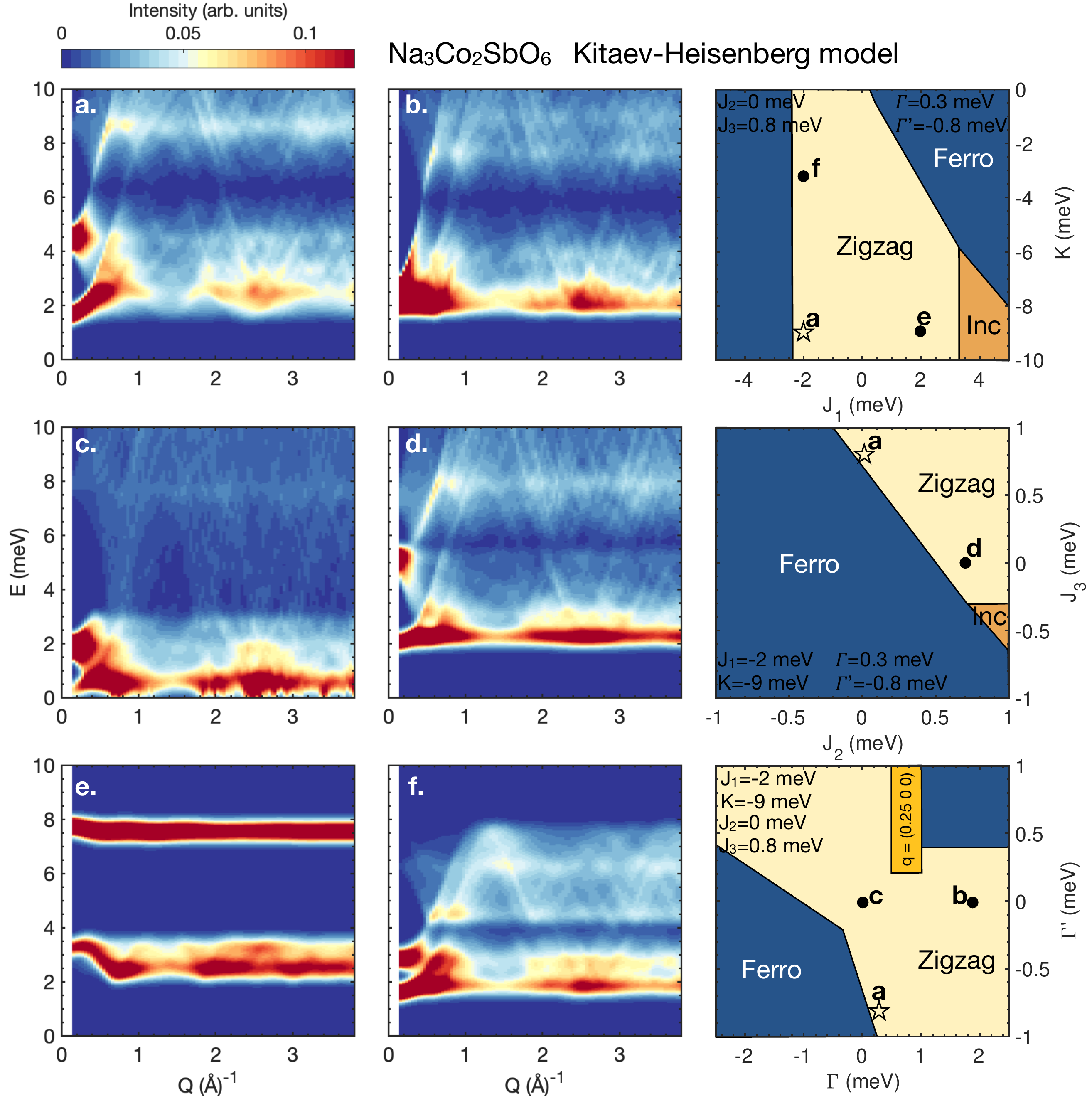}
 \caption{\label{fig:KHparam_Sb} Dependence of the spin-wave calculation with respect to variation of the six parameters involved in the K-H model for Na$_3$Co$_2$SbO$_6$. The panel $(a)$ corresponds to the chosen solution, that best reproduces the experimental data. Each panel ($(a)$-$(f)$) corresponds to a set of parameters, which are represented in the phase diagram (right panels). The blue, yellow and orange regions correspond to ferromagnetic, zig-zag and incommensurate magnetic structures, respectively.}
\end{figure*} 

Several sets of parameters were tested to best reproduce the experimental data. The results obtained for Na$_2$Co$_2$TeO$_6$ and Na$_3$Co$_2$SbO$_6$ are shown in Figures \ref{fig:KHparam_Te} and \ref{fig:KHparam_Sb}, respectively. The tested sets of parameters are represented in the phase diagrams displayed in the right panels and are all compatible with a zigzag magnetic order. In order to be consistent with previous theoretical calculations \cite{Liu2018,Sano2018,Liu2020}, the Kitaev term $K$ is ferromagnetic. As inferred from the phase diagrams, the zigzag order weakly depends on the strength of the Kitaev and off-diagonal terms $K$, $\Gamma$ and $\Gamma'$. The effect of the off-diagonal couplings on the excitation spectra was tested and shown in Figures \ref{fig:KHparam_Te} $(b)$ and  \ref{fig:KHparam_Sb} $(b)$ and $(c)$. The results show that the $\Gamma$ term introduces a gap in energy while $\Gamma'$ modifies the shape of the lower mode: in the Sb case, its absence causes the magnetic excitation around 2 meV to be flat in momentum and energy. In the case of the Te compound, this term was found to be negligible. The magnitude of the $K$ coupling, which determines the overall energy range of the magnetic excitations was also adjusted (Fig.\ref{fig:KHparam_Te} $(c)$ and  \ref{fig:KHparam_Sb} $(f)$). In particular, in the Te case, the ratio between the 3 and 7 meV flat modes is obtained through the Kitaev interaction. In the Sb case, a subtle combination of all parameters allows to reproduce the strong intensity at low Q, the flat mode around 3 meV with a  kink at 2 \AA$^{-1}$ and a dispersive feature with maximum around 9 eV.

While the zigzag groundstate is stable over a large range in $K$, $\Gamma$ and $\Gamma'$, it strongly depends on the interplay between the Heisenberg terms $J_1$, $J_2$ and $J_3$. Therefore, both the magnitude and signs of $J_1$ were tested, as shown in Figures \ref{fig:KHparam_Te} $(e)$ and $(f)$ and \ref{fig:KHparam_Sb} $(e)$. It can be noted that an antiferromagnetic $J_1$ coupling does not allow to reproduce the data, and that $J_1$ has to be ferromagnetic. In the case of the Te compound, we found that $J_1 \sim 0$ best reproduces the data. The calculated phase diagrams also show that antiferromagnetic $J_2$ and $J_3$ terms are necessary to stabilise a zigzag groundstate. Following these results, Figures \ref{fig:KHparam_Te} $(a)$, \ref{fig:KHparam_Te} $(d)$, \ref{fig:KHparam_Sb} $(a)$ and $(d)$ show that $J_3 \geq J_2$ gives a better agreement with the data.

\subsubsection{Heisenberg XXZ model}

\begin{figure*}
 \includegraphics[scale=0.48]{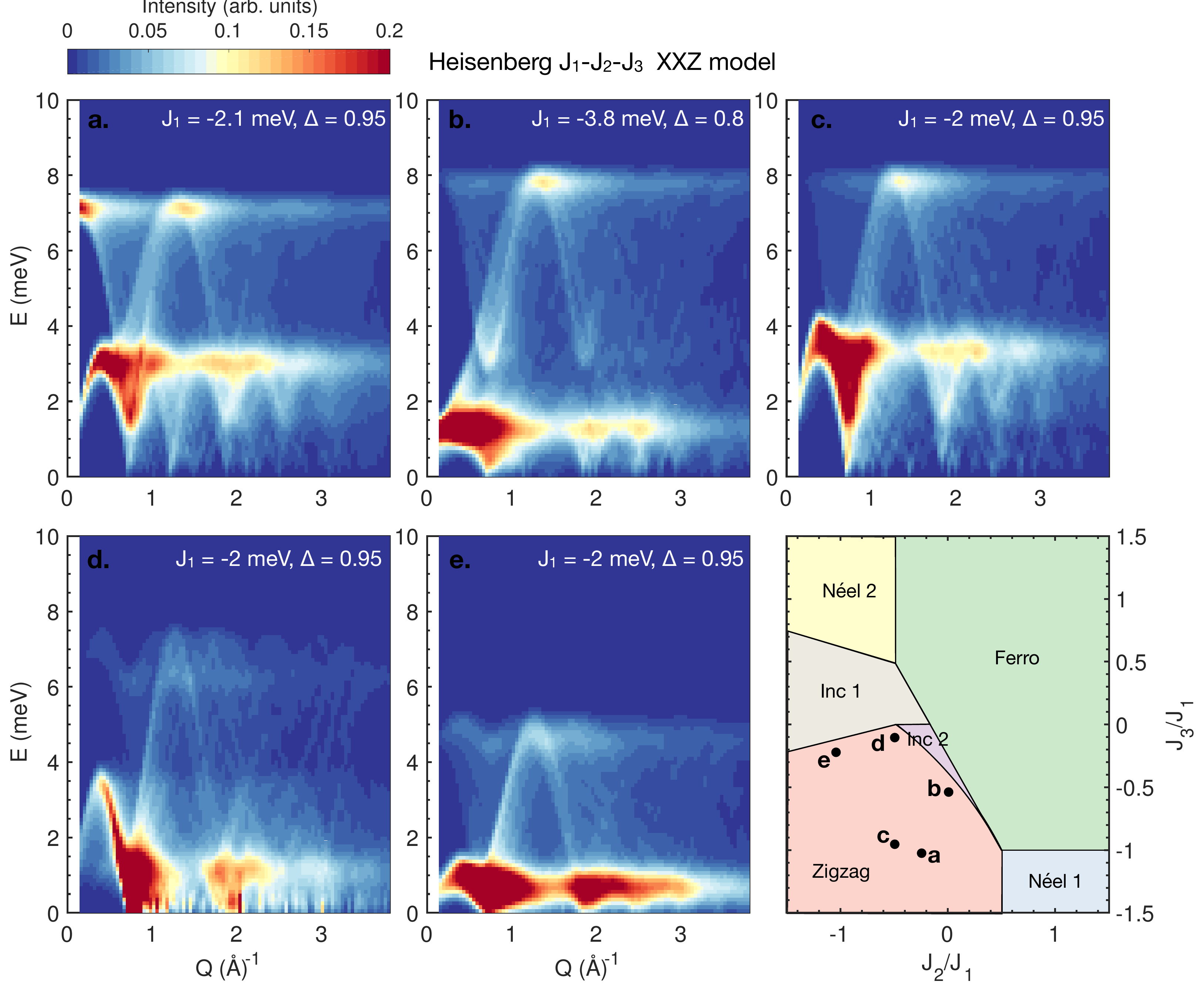}
 \caption{\label{fig:XXZparam} Dependence of the spin-wave calculations with respect to variation of the first, second and third nearest-neighbour couplings involved in the Heisenberg XXZ model, for a honeycomb lattice system. The sets of parameters are represented in the phase diagram (bottom right panel). No interplane coupling was considered in these calculations. Panels $(a)$ and $(b)$ are the best-fit solutions for Na$_2$Co$_2$TeO$_6$ and Na$_3$Co$_2$SbO$_6$, respectively. The colors in the phase diagrams refer to the different possible magnetic groundstates identified in \cite{Lefrancois2016}}
\end{figure*}

\begin{figure*}
 \includegraphics[scale=0.45]{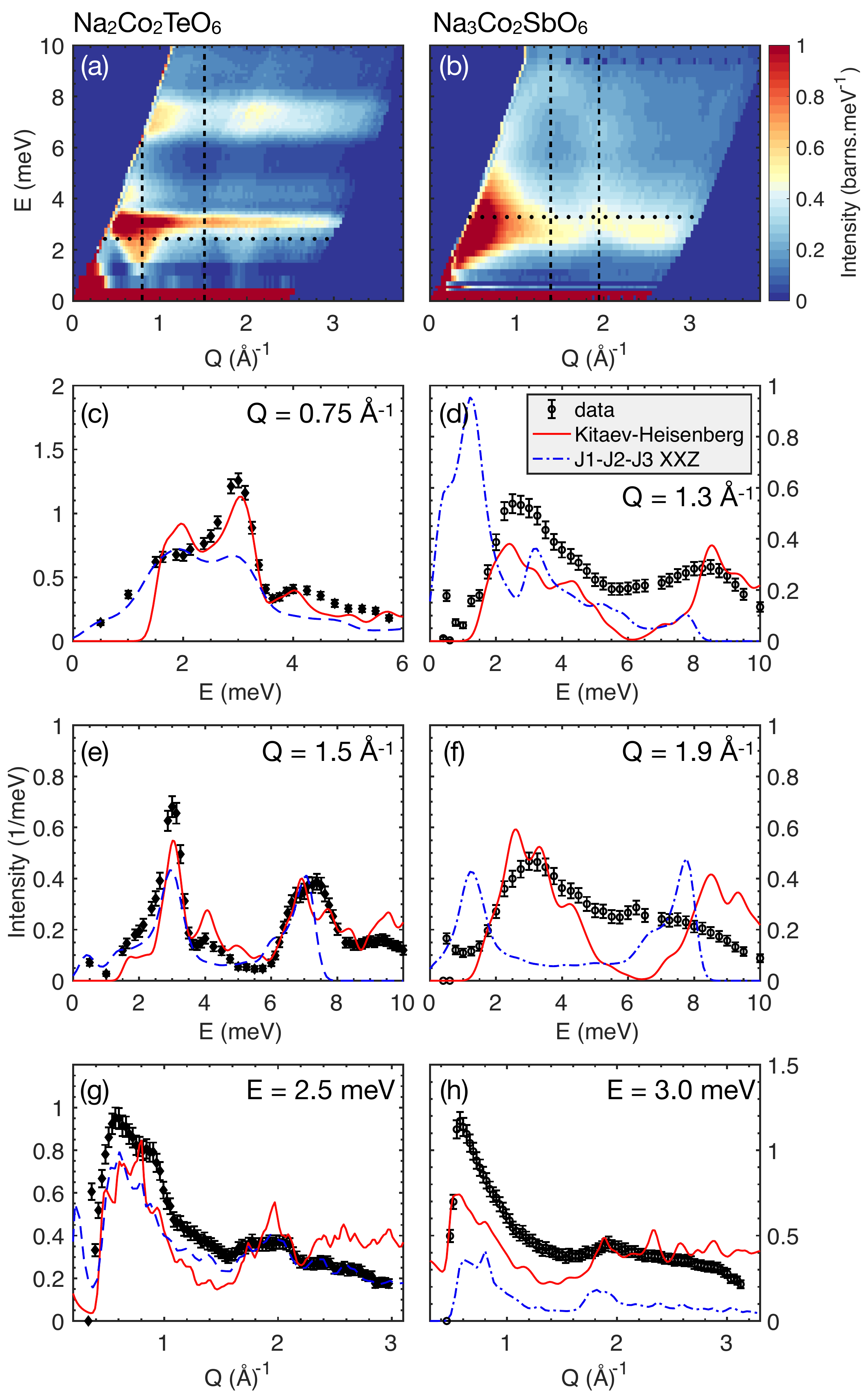}
 \caption{ \label{fig:cuts} $(a)$-$(b)$ S(\textbf{Q},$\omega$) at T = 1.5 K measured on MACS for Na$_2$Co$_2$TeO$_6$ and Na$_3$Co$_2$SbO$_6$, respectively. Comparison of the Kitaev-Heisenberg model and the $J_1$-$J_2$-$J_3$ model with the data through constant-Q cuts ($(c)$-$(f)$) and constant-energy cuts ($(g)$-$(h)$).}
\end{figure*}

The experimental spectra were also compared against a Heisenberg $XXZ$ model comprising first, second and third nearest-neighbour couplings ($J_1$-$J_2$-$J_3$ model) \cite{Nair2018}:

\begin{eqnarray*}
H_{XXZ} = & \sum_{n=1}^3 J_n  \sum_{i,j} (S_i^xS_j^x + S_i^yS_j^y +  \Delta S_i^zS_j^z)
\end{eqnarray*}

where $i$ and $j$ run over the first, second and third neighbour pairs, as shown in Figure 1, and $\Delta$ is comprised between 0 (XY anisotropy) and 1 (Heisenberg).
 Several sets of parameters, which are compatible with a zigzag magnetic order, were tested and are presented in Figure \ref{fig:XXZparam}. Following these results, we found that the best fit to the data are given by models \ref{fig:XXZparam} $(a)$ and \ref{fig:XXZparam} $(b)$ for the Te and Sb compounds, respectively, and are also summarized in Figure \ref{fig:macs}. No interplane interaction $J_4$ was considered in these calculations (and was found to be negligible, of the order of 0.01 meV for the Te compound \cite{Lefrancois2016}). 
These calculations however show a poorer agreement with the data. This is further highlighted in constant-Q cuts and constant energy cuts displayed in Figure \ref{fig:cuts}, showing a comparison between the K-H model, the Heisenberg XXZ model and the experimental data. In particular, the $J_1$-$J_2$-$J_3$ model fails to capture several features, such as the 4.5 meV flat mode and intensity above 7 meV for the Te compound, and the intensity around Q~=~2~\AA$^{-1}$, as well as the energy gap for the Sb compound (Figure \ref{fig:macs}). These features could only be reproduced by adding the bond-dependent interactions $K$, $\Gamma$ and $\Gamma'$. 
As pointed out in a recent theoretical analysis on Na$_3$Co$_2$SbO$_6$, a Heisenberg XXZ model would entail a large distortion of the octahedra and does not account for the bond-dependent interactions from the pseudospin-1/2 picture in cobaltates \cite{Liu2020}.

\subsection{Temperature dependence}

\begin{figure*}
 \includegraphics[scale=0.45]{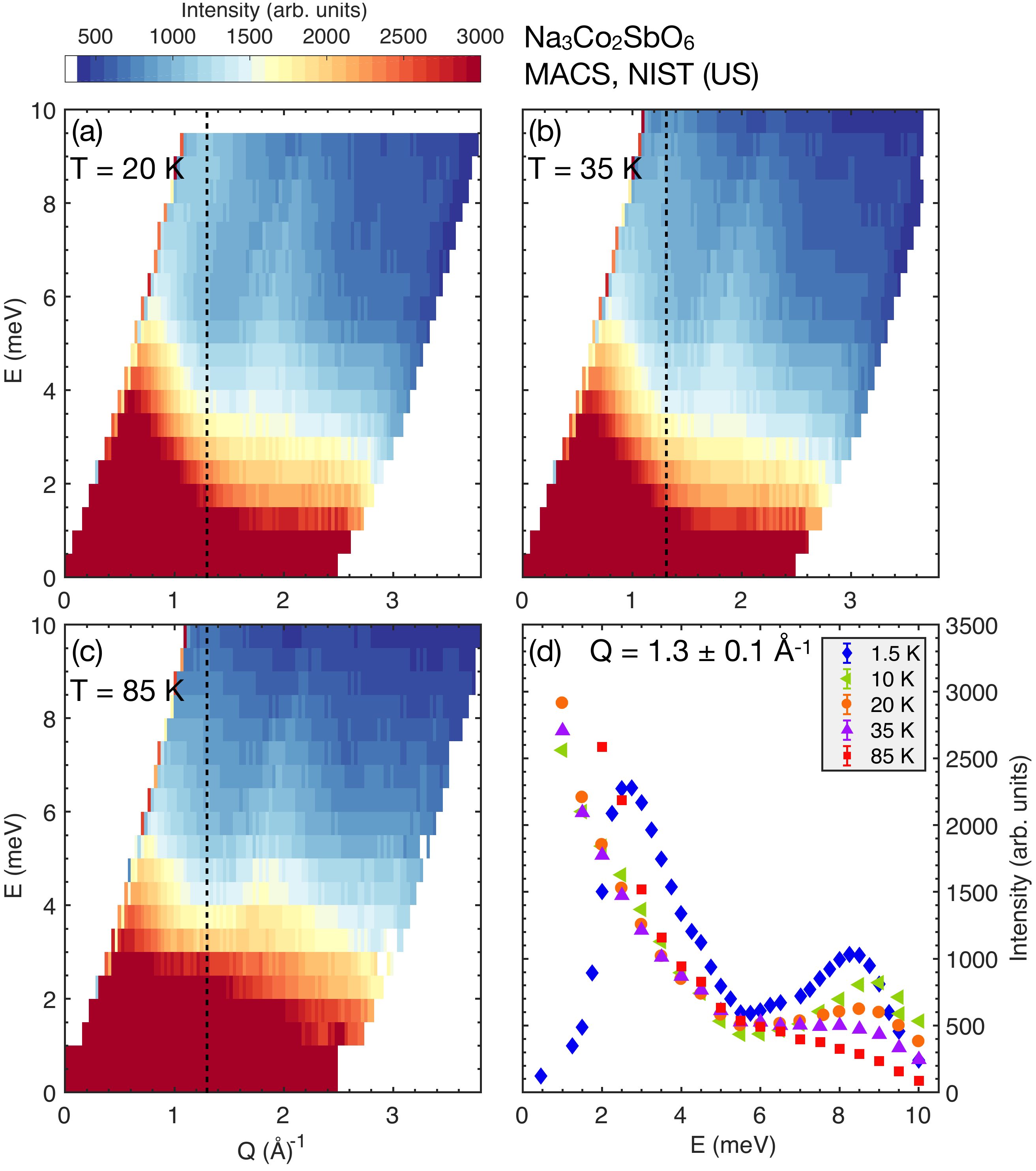}
 \caption{\label{fig:evol_temp} Temperature dependence of the magnetic excitations in Na$_3$Co$_2$SbO$_6$.}
\end{figure*}

The temperature dependence of the magnetic excitations in Na$_3$Co$_2$SbO$_6$ was measured on MACS and is shown in Fig. \ref{fig:evol_temp}. The spectrum shows persisting excitations up to 10 meV well above the N\'eel temperature. In particular, Figure \ref{fig:evol_temp} $(d)$ displays the momentum-integrated data for several temperatures and highlights a well-defined mode around 8 meV which persists up to $\sim$ 5 T$_N$. This is indicative of short-range correlations surviving within the honeycomb plane which can be due to the low-dimensionality of the system. While this feature has been reported in numerous low-dimensional magnets \cite{Songvilay2018,Songvilay2017}, such high-temperature feature was also attributed to the possible manifestation of fractional excitations in the powder sample of $\alpha$-RuCl$_3$ \cite{Banerjee2016}, arising from its proximity to a Kitaev quantum spin liquid phase. 

\section{Discussion}

Remarkably, in both cases, the leading term of the Hamiltonian is the Kitaev interaction, which is ferromagnetic, in agreement with the predictions for Co$^{2+}$ \cite{Liu2018,Sano2018}. In the Te compound, the second strongest term is the bond-directional $\Gamma$ interaction, which sets the energy gap of the excitation, whereas the first-neighbour Heisenberg interaction is close to zero. This is a situation predicted by Liu {\it et al} \cite{Liu2018} and Sano {\it et al} \cite{Sano2018} for $d^7$ pseudo-spin-1/2 where the Heisenberg interaction can vanish leaving a leading Kitaev interaction, placing the system in proximity to a spin liquid state. In Na$_2$Co$_2$TeO$_6$ however, additional weaker terms drive the system away from the spin liquid groundstate. In particular, as predicted in \cite{Liu2018,Sano2018}, for a ferromagnetic $K$ coupling, next-nearest-neighbour terms $J_2$ and $J_3$ are necessary to stabilize the zigzag magnetic order.

In the Sb case, the nearest-neighbour interaction is significant, although still five times smaller than the Kitaev term, and is ferromagnetic. It should also be noted that the Sb compound is close to an instability due to the proximity of a ferromagnetic phase (Figure \ref{fig:diagphase} right panels), in agreement with recent theoretical calculations \cite{Liu2020}. This may explain the very different spectra measured in the Sb case compared to the Te variant. The difference between the Sb and Te compounds is however rather intriguing as they display similar magnetic properties and magnetic structures. This may be attributed to structural distortions away from the perfect octahedral environment (as illustrated in Fig. \ref{fig:crystal_field} $(b)$) and to further-neighbour couplings. These distortions also account for the additional off-diagonal interaction $\Gamma'$, which affects strongly the spectrum in the Sb case while it remains negligible in the Te compound. Further structural study is however needed to understand these differences, as disparate results were reported for the magnetic structure and the ordered moment of Na$_3$Co$_2$SbO$_6$ \cite{Stratan2019,Wong2016,Yao2019}. In this compound, complementary neutron and x-ray diffraction measurements were performed and show a slightly incommensurate magnetic order, probably arising from sodium vacancies (see Supplemental Information). A more precise refinement of the magnetic order and moments orientation in this compound would therefore help to determine a more accurate set of exchange parameters.

These results underline the potential realization of dominant ferromagnetic Kitaev interactions in cobaltates, despite featuring a weaker spin-orbit coupling than the $4d$ and $5d$ counterparts. While our analysis shows the relevance of the six exchange parameters in determining the magnetic structure, along with reproducing the main features in the excitation spectra, the different spin-wave models were further compared to the data through constant-Q cuts and highlight some discrepancies between the K-H model and the data above 6 meV. It should be noted that our analysis was performed using a classical model, while we could expect quantum fluctuations to renormalise the excitation spectra or to cause energy and momentum broadening of the intensity, as reported in other low-dimensional frustrated magnets \cite{Mourigal2013,Zhitomirsky2013,Songvilay2018,Ma2016,Ito2017}. Furthermore, a recent analysis of $\alpha$-RuCl$_3$ has shown that the presence of off-diagonal terms ($\Gamma$ and $\Gamma'$) causing the spins to tilt out of the honeycomb plane, could induce magnon decay \cite{Winter2017,Smit2020,Maksimov2020}. This effect would be expected at twice the energy of the main magnon branch and therefore could explain the discrepancies above 6 meV in both compounds, where the scattering is broad in energy and momentum. 

The synthesis of small single crystalline samples of Na$_2$Co$_2$TeO$_6$ and Na$_3$Co$_2$SbO$_6$ have been recently reported \cite{Yao2019, Yan2019}. Further spectroscopic measurements in single crystals are therefore necessary to confirm the present models obtained from the limited set of information provided by the powder averaged data, as well as determine the presence of quantum effects in these compounds. Note that the compound BaCo$_2$(PO$_4$)$_2$ exhibits an excitation spectrum similar to Na$_3$Co$_2$SbO$_6$, hosts a helical magnetic order and was modeled using a XXZ Heisenberg Hamiltonian \cite{Nair2018}. The relevance of this model implies a large trigonal distortion of the Co$^{2+}$ octahedral environment \cite{Liu2020}. While these materials (along with the isostructural BaCo$_2$(AsO$_4$)$_2$) have been investigated since several decades \cite{Regnault1977,Regnault2018,Martin2012,Zhong2020}, it seems timely to revisit their magnetic properties and excitations in the context of bond-dependent Kitaev interactions.

\section{Conclusion }
In conclusion, we report a neutron spectroscopy study of the powder honeycomb cobaltates Na$_2$Co$_2$TeO$_6$ and Na$_3$Co$_2$SbO$_6$. Using linear spin wave calculations, we show that the magnetic excitations in both compounds can be modeled using a Kitaev-Heisenberg Hamiltonian with additional off-diagonal bond-directional interactions and long-range Heisenberg interactions. The calculations highlight dominant bond-dependent Kitaev interactions and provide direct evidence for the possibility of stabilizing Kitaev physics in cobaltates despite their weak spin-orbit coupling. Our results show the possibility for extending the search for new Kitaev spin liquid candidates to $3d$ metal transition compounds.

\begin{acknowledgments}
We thank G. Khaliullin for helpful discussion. We acknowledge funding from the EPSRC and the STFC. We acknowledge the support of the National Institute of Standards and Technology, U.S. Department of Commerce, in providing the neutron research facilities used in this work. Access to MACS was provided by the Center for High Resolution Neutron Scattering, a partnership between the National Institute of Standards and Technology and the National Science Foundation under Agreement No. DMR-1508249.
\end{acknowledgments}

\bibliography{Co-honeycomb}

\end{document}


\title{Supplemental Material for ``Kitaev interactions in the Co honeycomb antiferromagnets Na$_3$Co$_2$SbO$_6$ and Na$_2$Co$_2$TeO$_6$''}

\author{M. Songvilay}
\affiliation{School of Physics and Astronomy and Centre for Science at Extreme Conditions, University of Edinburgh, Edinburgh EH9 3FD, UK}
\affiliation{Institut N\'eel, CNRS \& Univ. Grenoble Alpes, 38000 Grenoble, France}
\author{J. Robert}
\affiliation{Institut N\'eel, CNRS \& Univ. Grenoble Alpes, 38000 Grenoble, France}
\author{S. Petit}
\affiliation{Laboratoire L\'{e}on Brillouin, CEA-CNRS UMR 12, 91191 Gif-Sur-Yvette Cedex, France}
\author{J. A. Rodriguez-Rivera}
\affiliation{NIST Center for Neutron Research, National Institute of Standards and Technology, 100 Bureau Drive, Gaithersburg, Maryland,20899, USA}
\affiliation{Department of Materials Science, University of Maryland, College Park, Maryland 20742, USA}
\author{W. D. Ratcliff}
\affiliation{NIST Center for Neutron Research, National Institute of Standards and Technology, 100 Bureau Drive, Gaithersburg, Maryland, 20899, USA}
\author{F. Damay}
\affiliation{Laboratoire L\'{e}on Brillouin, CEA-CNRS UMR 12, 91191 Gif-Sur-Yvette Cedex, France}
\author{V. Bal\'edent}
\affiliation{Laboratoire de Physique des Solides, CNRS, Universit\'e Paris-Sud, Universit\'e Paris-Saclay, 91405 Orsay Cedex, France}
\author{M. Jim\'enez-Ruiz}
\affiliation{Institut Laue-Langevin, 71 avenue des Martyrs, 38000 Grenoble, France}
\author{P. Lejay}
\affiliation{Institut N\'eel, CNRS \& Univ. Grenoble Alpes, 38000 Grenoble, France}
\author{E. Pachoud}
\affiliation{Institut N\'eel, CNRS \& Univ. Grenoble Alpes, 38000 Grenoble, France}
\author{A. Hadj-Azzem}
\affiliation{Institut N\'eel, CNRS \& Univ. Grenoble Alpes, 38000 Grenoble, France}
\author{V. Simonet}
\affiliation{Institut N\'eel, CNRS \& Univ. Grenoble Alpes, 38000 Grenoble, France}
\author{C. Stock}
\affiliation{School of Physics and Astronomy and Centre for Science at Extreme Conditions, University of Edinburgh, Edinburgh EH9 3FD, UK}
\date{\today}

\maketitle

\section{Diffraction studies on Na$_3$Co$_2$SbO$_6$}

Additional neutron powder diffraction on Na$_3$Co$_2$SbO$_6$ was carried out using the G4.1 diffractometer (LLB, Saclay) and powder x-ray diffraction was performed on the same compound at the Laboratoire de Physique des Solides (Orsay). 

\begin{figure}
 \includegraphics[scale=0.26]{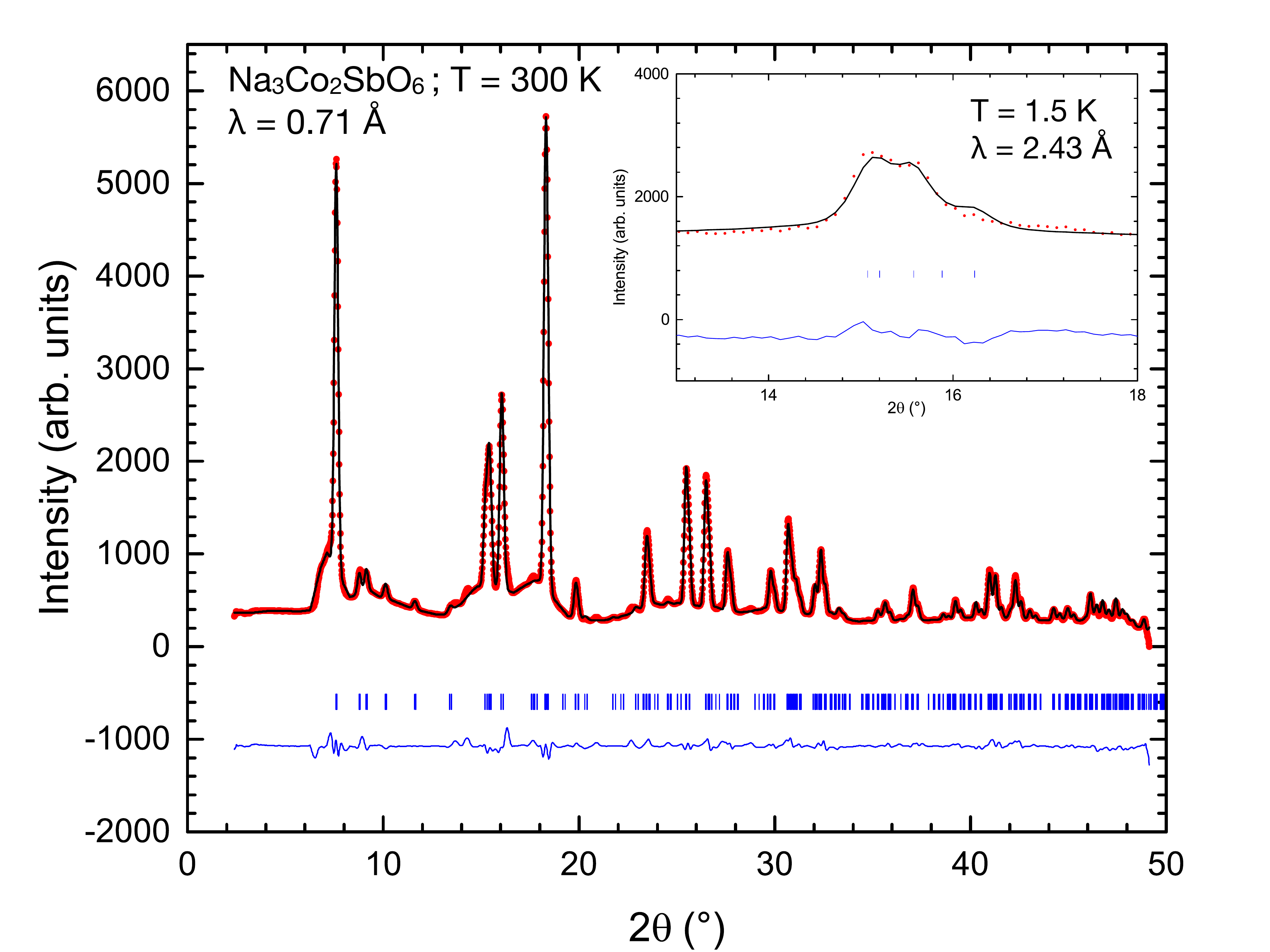}
 \caption{\label{fig:diff} X-ray diffraction data of Na$_3$Co$_2$SbO$_6$ measured at T = 300 K. The insert shows a zoom on magnetic Bragg peaks measured by neutron diffraction on the G41 spectrometer at T = 1.5 K. The red dots represent the experimental data, the black line is the calculated intensity, the blue line is the experimental intensity subtracted from the calculated intensity and the blue ticks are the calculated Bragg positions.}
\end{figure}

A Rietveld refinement of the X-ray diffractogram of Na$_3$Co$_2$SbO$_6$ is shown in Fig. \ref{fig:diff} and points to some deficiency on the sodium and antimony sites, with a substantial thermal parameter (the refined structural parameters are summarised in Table \ref{tab:fullprof}). These results thus highlight potential structural disorder in this compound. Neutron powder diffraction at low temperature revealed magnetic Bragg peaks which could be fitted with a slightly incommensurate propagation vector, using a Le Bail fit (see insert of Fig. \ref{fig:diff}). The fitted propagation vector is \textbf{k}~=~(0.48(3) 0.47(3) 0). Deficiencies on the sodium and antimony sites may lead to structural modulation, such as stacking faults or buckling of the cobalt layers, as observed in the triangular lattice compound $\beta$-NaMnO$_2$ \cite{Orlandi2018}, which may account for the slightly incommensurate magnetic structure found in this Sb compound. Furthermore, antimony vacancies might induce the cobalt ions to be in a Co$^{3+}$ state, which is non-magnetic. However, our analysis using the zeroth moment sum rule (detailed in the main article) shows that the total spectral weight contained in the inelastic scattering corresponds, within error, to the expected value for a j$_{eff}$ = 1/2 system. We therefore expect that even a small amount of Co$^{3+}$ in the system should not affect the excitation spectrum.

\begin{table}[t]
\caption{\label{tab:fullprof} Structural refinement of Na$_3$Co$_2$SbO$_6$ in the $C2/m$ space group at 300 K with $a$~=~5.366(7)~\AA, $b$~=~9.279(1)~\AA , $c$~=~5.655(5)~\AA ~and $\beta$ = 108.56(1)$^{\circ}$ (RF factor = 10.1)}
\begin{ruledtabular}
\begin{tabular}{lllllll}
 Atom & Wyckoff & $x$ & $y$ &$z$ & Occ. & $U_{\rm iso}$\\
 \hline
Co & $4g$ & 0 & 0.663(7) & 0 & 1 & 0.26\\
Sb & $2a$ & 0 & 0 & 0 & 0.788(2) & 1.19(5)\\
O(1) & $8j$ & 0.309(2) & 0.353(2) & 0.808(2) & 1& 0.21 \\
O(2) & $4i$ & 0.288(8) & 1/2 & 0.165(2) & 1 & 0.21(10)\\	
Na(1) & $2d$ & 0 & 1/2 & 1/2 & 0.93(2) & 1.27(11)\\
Na(2) & $4h$ & 1/2 & 0.334(2) & 1/2 & 0.89(1) & 1.27(11)\\
\end{tabular}
\end{ruledtabular}
\end{table}

\section{Neutron inelastic complementary measurements}

\begin{figure}
 \includegraphics[scale=0.32]{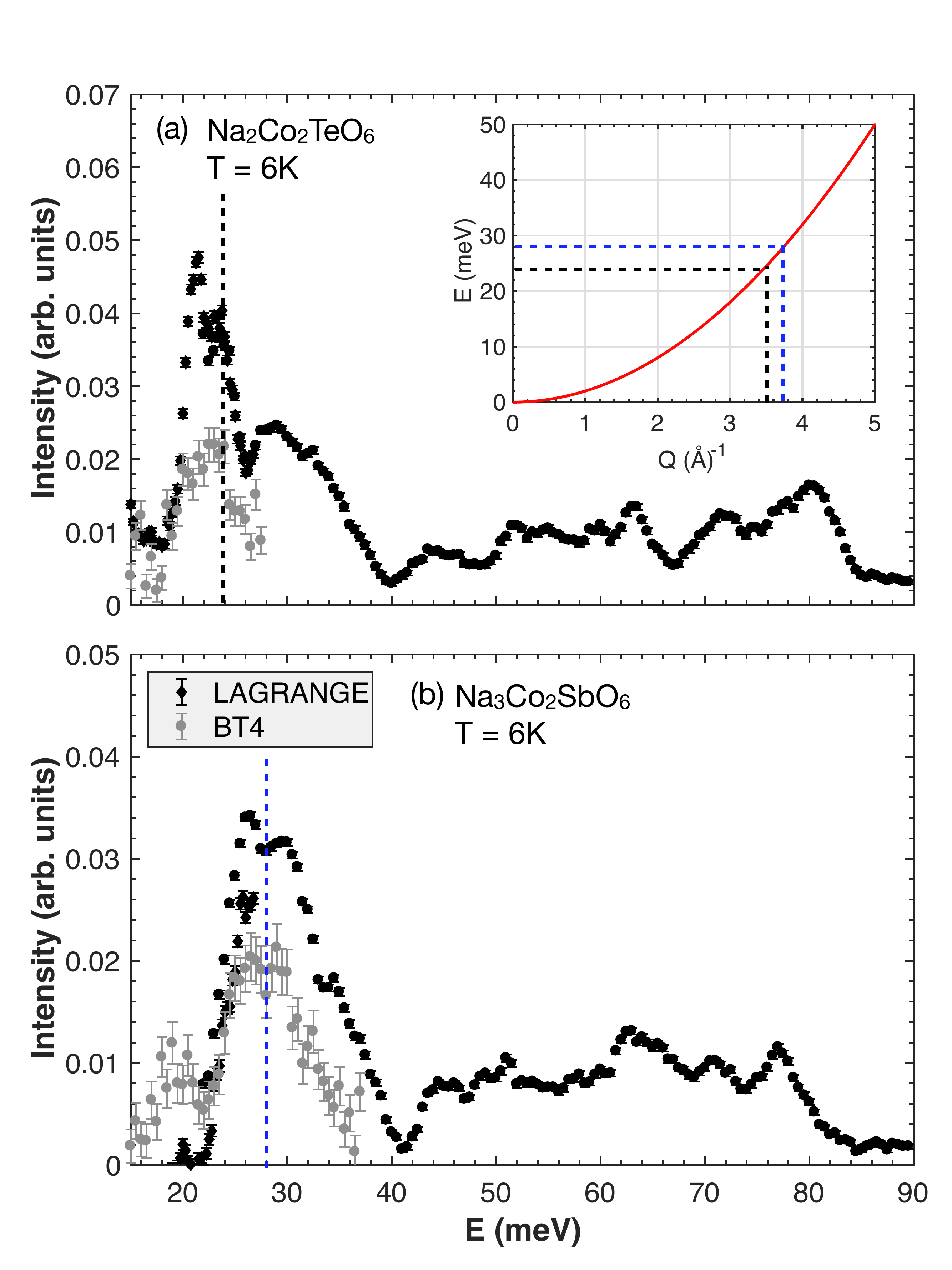}
 \caption{\label{fig:IN1} Neutron inelastic data measured at high energy on the LAGRANGE spectrometer (ILL, France) at T = 6 K for $(a)$ Na$_2$Co$_2$TeO$_6$ and $(b)$ Na$_3$Co$_2$SbO$_6$. The insert shows the energy and momentum coverage of the LAGRANGE detector. The data shown in gray corresponds to constant-Q scans at Q~=~2.5 \AA$^{-1}$ and Q~=~2.75 \AA$^{-1}$, respectively, measured on BT4. The intensity was multiplied by a factor in order to use the same intensity scale.}
\end{figure} 

Complementary measurements were carried out up to 150 meV on the LAGRANGE spectrometer at the ILL (France), using the Si111, Si311 and Cu220 monochromators to access different energy ranges between 0 and 150 meV. In this experiment, a single detector records the intensity scattered over a large solid angle \cite{Ivanov2014}, thanks to a focussing large graphite analyser, and is used to cover a wide range in energy transfer, following kinematic conditions, as shown in the energy and Q dependence in the insert of Figure \ref{fig:IN1} $(a)$.

Figure \ref{fig:IN1} shows the excitations spectrum measured on the LAGRANGE spectrometer at low temperature, in an energy range between 15 and 90 meV for both compounds. The insert shows the energy and momentum coverage of the detector, following kinematic conditions. A comparison with constant-Q scans measured on BT4 is also shown in gray, at Q~=~2.5 \AA$^{-1}$ for the Te compound and Q~=~2.75 \AA$^{-1}$ for the Sb compound. As displayed on figures \ref{fig:IN1} $(a)$ and $(b)$, these constant-Q cuts are in agreement with peaks observed on LAGRANGE around 24 and 28 meV, respectively. The BT4 data (Figure 4 $(a)$) discussed in the main manuscript  follows the momentum dependence of the expected magnetic form factor for Co$^{2+}$ and was therefore assigned to excitations between the $j_{\rm eff}$ = 1/2 and $j_{\rm eff}$ = 3/2 manifolds. It can be inferred from this comparison that the signal observed below 30 meV on LAGRANGE corresponds to magnetic scattering. Above this energy, the detector coverage corresponds to a high Q region (above 4 \AA$^{-1}$ according to the insert of figure \ref{fig:IN1} $(a)$); the signal may  thus be associated to phonons contribution, which intensity should grow as $| \mathbf{Q} | ^2$.

\bibliography{Co-honeycomb}